\documentclass{aa}  
\usepackage{color}

\newcommand{\nb}{n_\mathrm{B_{rad}}}
\newcommand{\Alfven}{Alfv\'{e}n}
\usepackage{graphicx}
\usepackage{txfonts}
\begin{document} 

   \title{Standing torsional Alfv\'{e}n waves as the source of the rotational period variation in magnetic early-type stars}
   
   \titlerunning{Standing torsional Alfv\'{e}n waves}
   \author{Koh Takahashi\inst{1,2},
            Norbert Langer\inst{3,4}
    }
   \institute{
            National Observatory of Japan, National Institutes for Natural Science, 2-21-1 Osawa, Mitaka, Tokyo 181-8588, Japan\\
            \email{koh.takahashi@nao.ac.jp}
         \and
            Astronomical Institute, Tohoku University, 6-3 Aramaki Aza-Aoba, Aoba, Sendai, Miyagi 980-8578, Japan
         \and
            Argelander-Institut f\"{u}r Astronomie, Universit\"{a}t Bonn, Auf dem H\"{u}gel 71, 53121 Bonn, Germany
         \and
            Max-Planck-Institut f\"{u}r Radioastronomie, Auf dem H\"{u}gel 69, 53121 Bonn, Germany
    }
   \date{Date:--}

% \abstract{}{}{}{}{} 
% 5 {} token are mandatory
 
  \abstract
   {The influence of magnetic fields on stellar evolution remains unresolved. It has been proposed that if there is a large-scale magnetic field in the stellar interior, torsional \Alfven{} waves could arise, efficiently transporting angular momentum. In fact, the observed variations in the rotation periods of some magnetic stars may be attributed to these torsional \Alfven{} waves' standing waves.}
  % aims heading (mandatory)_
   {To demonstrate the existence of torsional \Alfven{} waves through modeling of the rotational period variations.}
   {We conduct an eigenmode analysis of standing \Alfven{} waves based on one-dimensional magnetohydrodynamic equations. The internal magnetic field structures are parametrically represented to treat poloidal fields with different degrees of central/surface concentration. The obtained frequencies are compared with the observed frequencies of the rotational period variations, thereby constraining the internal magnetic field structures.}
   {The 67.6 years exhibited by CU Vir is reproduced for surface-concentrated magnetic field structures. The rotational period variations of all ten magnetic stars analyzed in this study are inconsistent with a centrally concentrated magnetic field.}
   {Torsional \Alfven{} waves can reproduce the observations of rotational period variations. The large-scale magnetic fields within magnetic stars would be concentrated on the surface.}


   \maketitle
%-------------------------------------------------------------------
\section{Introduction}

About 10\% of early-type (O-, B-, and A-type) stars are known to possess surface magnetic fields of the order of 100--10,000 G \citep{Babcock58, Landstreet92, Wade+14, Keszthelyi23}.
In contrast to low-mass stars including the sun, the magnetic fields in these massive stars have a large-scale, time-stable structure. These properties support the idea that magnetic early-type stars possess a "fossil field", i.e., a magnetic field in mechanical equilibrium with the stellar plasma \citep{Braithwaite17}.
These stars are discussed as progenitors of compact objects with strong magnetic fields \citep[e.g.,][]{Ferrario06_Pulsar_Origin}.

Several processes have been proposed through which magnetic fields may influence stellar evolution.
If the magnetic energy density exceeds the internal energy density, magnetic pressure and tension can change the pressure and density structure of the star \citep{Feiden&Chaboyer12, Feiden&Chaboyer13, Feiden&Chaboyer14}. 
Even if it were not so strong, it could affect convection, meridional circulation, and rotationally induced flows with relatively small energy densities \citep[e.g.,][]{Petermann+15}.
Several studies investigated the effect of magnetic stress onto the rotational evolution.
External magnetic fields may exert torques on the stellar surface, braking its rotation \citep{Weber&Davis67, Meynet+11, Petit+17, Georgy+17, Keszthelyi+19}. Similarly, the magnetic field inside the star likely causes angular momentum transport and affects the rotation distribution \citep{Spruit99, Spruit02, Heger+05, Wheeler15, Fuller19, Griffiths22}.

There also have been attempts to calculate the temporal evolution of the global magnetic field within magnetic stars \citep{Potter12,Takahashi&Langer21}. The latter devised a formulation that satisfies conservation laws.
If there is a poloidal (component within the meridional plane) magnetic field within a rotating star, the differential rotation stretches it and induces a toroidal (component winding around the magnetic axis) component. As a result, a magnetic torque is induced which drives waves of angular momentum through the poloidal magnetic field: torsional \Alfven{} waves.
\citet{Takahashi&Langer21} discovered that the propagation of the dissipating torsional \Alfven{} waves redistributes angular momentum efficiently inside the star, providing an explanation for the slow rotation of the cores of red giants \citep{Beck12, Mosser12, Deheuvels14}.

Therefore, the question of whether there is a large-scale magnetic field inside a star is crucial because it can affect the overall stellar evolution. While observational evidence has verified the presence of a global magnetic field on the surface of magnetic stars, how this field is distributed and connected within the stellar interior remains unexplored. 
Asteroseismic analyses have suggested the presence of strong magnetic fields within the interiors of stars \citep{Fuller15, Stello16}, and more quantitative estimates have been made recently (for B-type main-sequence stars: \citealt{Lecoanet22}; for red giants: \citealt{Li22, Li23}). However, current magnetic field detections via asteroseismology are sensitive to only a narrow region within the stellar interior (the core surface), and the global structure remains largely unconstrained.

In this work, we focus on the intriguing phenomenon observed in some hot magnetic stars regarding their rotational periods to address the issue.
A- and B-type stars with magnetic fields are known to exhibit chemical peculiarities in their spectra, leading to their classification as Ap/Bp stars \citep{Maury97_HDMemorial, Lockyer06, Morgan33, Preston74}.
It was already noted in the early 1900s that these peculiar lines show time variation \citep{Lockyer06, Ludendorff06}, and later, its periodic nature with a period of 5.5 days was demonstrated for $\alpha^2$ CVn, known today as a typical Ap star \citep{Belopolsky13}.
Concurrently, a periodic variation in the stellar luminosity was also discovered, and it was shown that this period matches the spectroscopically determined period \citep{Guthnick&Prager14}.

After the detection of surface magnetic fields on 78 Vir \citep{Babcock47_detection}, the strength of the magnetic field \citep{Babcock47_variation, Babcock48}, line intensities \citep{Morgan31, Morgan33, Deutsch47}, and the luminosity \citep{Stibbs50} of CU Vir were all found to vary synchronously.
This variation of multiple quantities with a single period is currently explained with the Oblique Rotator Model, which assumes that the magnetic field symmetry axis is inclined to the rotation axis of the magnetic stars
%attributed to the effect of observing magnetic stars from different viewing angles due to their rotation 
(\citealt{Babcock49_Gentlemen, Stibbs50, Deutsch54_FirstObliqueRotatorModel, Deutsch58_FirstTomography, Wolff83_monogram}).

Due to this historical background, there exist long-term, high-precision observations of the rotation periods for some magnetic stars. 
Such an extended monitoring allows observers to discover changes of the rotation periods occurring over timescales of several decades.
Such changes were initially detected through phase shifts in light curves (56 Ari in \citealt{Adelman91}; CU Vir in \citealt{Adelman92}). Subsequently, \citet{Pyper98} introduced the O-C analysis and accurately confirmed the phase shifts for CU Vir.
Entering the 2000s, not only stars with increasing rotation periods, which indicate a deceleration in rotation velocity \citep{Reiners00, Mikulasek08, Townsend10}, but also stars showing decreasing rotation periods, suggesting accelerating rotation \citep{Ozuyar18_V473Tau, Shultz19_V913Sco, Pyper20}, were reported.

The rotational period variations discussed above indicate either a steady deceleration or acceleration in rotation, which corresponds to the first time derivative of the rotation period. 
Even more exciting, for CU Vir and V901 Ori, the presence of both, deceleration and acceleration phases, were indicated in the long-term  evolution of the rotation period \citep{Mikulasek11}. Additionally, V913 Sco, indicating an acceleration, may also exhibit a non-zero second derivative of the rotation period \citep{Shultz19_V913Sco}.
The presence of accelerating and decelerating stars, and even the coexistence of both phenomena in a single star, may suggest that the cyclic variations in rotation periods are a universal phenomenon among hot magnetic stars \citep{Mikulasek22}.

Various explanations for the origin of rotational period variations have been proposed. 
\citet{Shore76} predicted that stars with magnetic fields, due to their non-spherical distortion, would undergo free or forced precession in cases of single or binary stars, respectively. According to the Oblique Rotator Model, a precessing magnetic star should alter the shape of its light curve, as well as shift phases of maximum and minimum light by changing the line-of-sight direction. Observational evidence has been obtained in several magnetic stars, including 56 Ari \citep{Adelman91, Ziznovsky00, Pyper21}. However, explaining stars like CU Vir, which exhibit rotational period variations while maintaining the shape of their light curves, is challenging to understand with precession alone \citep{Mikulasek11}.
Magnetic braking, resulting from the coupling between stellar winds and surface magnetic fields, can also cause rotational period variations. 
This process is expected in high-mass stars and may explain the deceleration of rotation in magnetic OB\,stars like $\sigma$ Ori E \citep{Townsend10}. 
However, its occurrence in relatively low-mass magnetic stars is uncertain, and explaining acceleration poses challenges.
Other hypotheses such as migrating spots and mass accretion have been considered \citep{Adelman92, Pyper98}, but whether these scenarios can explain the deceleration or acceleration observed over decades-long timescales remains uncertain.

Although the mechanism of rotational period variation is not yet understood, the global magnetic field of the stellar interior may be of relevance.
\citet{Stepien98} speculated that the coupling between the magnetic field and the meridional circulation in the stellar envelope may be responsible for the rotational period variation.
\citet{Krticka+17} further elaborated on this idea and attempted to explain the rotational period variation by standing torsional \Alfven{} waves. They assumed a goblet-like poloidal magnetic field penetrating the stellar interior, and they considered the torsional \Alfven{} waves by solving a one-dimensional induction equation.
Giving the modulating rotation as a boundary condition, they scanned the frequencies corresponding to the standing waves.
They found that the fundamental mode frequency corresponds well to the frequency of the rotational period variation observed for CU Vir.
Although their model could not explain V901 Ori, they demonstrated that standing \Alfven{} waves have the potential to explain rotational period variation.

In this study, the properties of standing \Alfven{} waves in magnetic stars are investigated in more detail, based on the method developed by \citet{Takahashi&Langer21}.
This method decomposes the large-scale magnetic field inside a star into poloidal and toroidal components, each of which can have an arbitrary radial dependence. It allows us to investigate the dependence of the variation period on the poloidal field geometry, beyond those considered in \citet{Krticka+17}.

The magnetic field structures considered here are still very simplified, and not capable of fully capturing the complex structures found in real stellar magnetospheres.
Therefore, a quantitative comparison between theory and observation may not be very meaningful.
Instead, we seek to qualitatively investigate which properties significantly affect the timescale of variability.
Ultimately, we aim to test the existence of torsional \Alfven{} waves and establish a picture of angular momentum transport in the stellar interior by large-scale magnetic fields through a comparison of predicted variations of the rotational period with observations.

In the next section, we compile the available observational constraints and explain our strategy for predicting the rotational period variation by internal \Alfven{} waves.
In Section \ref{sec:Results}, after analyzing the properties of standing torsional \Alfven{} waves using CU Vir as an example, we will derive constraints for the internal magnetic field structure of ten magnetic stars. 
In Section \ref{sec:Discussion}, we discuss the limitations of this model, its relation to previous research on the internal magnetic field structures of stars, the excitation mechanisms of torsional waves, and the prospects for future detection of rotational period variation.
We summarize concluding remarks in Section \ref{sec:Conclusion}.

\section{Method} \label{sec:Method}

\subsection{Magnetic stars that show rotational period variations}

\begin{table*}[]
    \caption{A list of magnetic stars that show rotational period variations.}
    \label{tab:objects}
    \centering
    \resizebox{\textwidth}{!}{
    \begin{tabular}{ccccccccccc}
        \hline
        Object &HD &$\log T_\mathrm{eff}$ &$\log L/L_\odot$ &$B_\mathrm{pole}$  &$P_\mathrm{rot}$  &$T_\mathrm{baseline}$  & $dP_\mathrm{rot}/dt$ &$M_\mathrm{ini}$   &$\tau$ & $\tau/\tau_\mathrm{MS}$ \\
        -      &-        &K   &-    &kG   &d     &yr                     & 10$^{-2}$ s yr$^{-1}$ &$M_\odot$          &Myr    & - \\
        \hline
        $\sigma$ Ori E &37479  &4.361$\pm$0.039\tablefootmark{a} &3.51$\pm$0.21\tablefootmark{a} & $7.55^{+0.25}_{-0.25}$\tablefootmark{b} &  1.191\tablefootmark{a} & 43 (1974--2017)\tablefootmark{c} &  9.71$\pm$0.09 \tablefootmark{c} & 8.5 & 4.33 & 0.1 \\
        V901 Ori       &37776  &4.342$\pm$0.019\tablefootmark{a} &3.30$\pm$0.15\tablefootmark{a} & $6.1^{+0.7}_{-0.7}$\tablefootmark{a}    &  1.539\tablefootmark{a} & 36 (1977--2013)\tablefootmark{d} &  35.3$\pm$0.9  \tablefootmark{c} & 7.5 & 0.00 & 0.0 \\
        V913 Sco       &142990 &4.255$\pm$0.012\tablefootmark{a} &2.93$\pm$0.13\tablefootmark{a} & $4.7^{+0.4}_{-0.4}$\tablefootmark{a}    &  0.979\tablefootmark{a} & 39 (1976--2015)\tablefootmark{e} & -58$\pm$1      \tablefootmark{e} & 5.5 & 27.6 & 0.3 \\
        56 Ari         &19832  &4.093$\pm$0.013\tablefootmark{a} &2.08$\pm$0.16\tablefootmark{a} & $2.7^{+0.6}_{-0.3}$\tablefootmark{a}    &  0.728\tablefootmark{a} & 48 (1952--2001)\tablefootmark{f} &  2.25          \tablefootmark{f} & 3.2 & 196  & 0.6 \\
        V343 Pup       &60431  &4.114$\pm$0.01\tablefootmark{g}  &2.00$\pm$0.032\tablefootmark{g}&   -                                     &  0.476\tablefootmark{g} & 35 (1986--2021)\tablefootmark{g} &  0.74$\pm$0.05 \tablefootmark{g} & 3.2 & 45.1 & 0.1 \\
        CU Vir         &124224 &4.089$\pm$0.003\tablefootmark{a} &1.93$\pm$0.01\tablefootmark{a} & $4.0^{+0.2}_{-0.2}$\tablefootmark{h}    &  0.521\tablefootmark{a} & 67 (1949--2016)\tablefootmark{c} &  8.817$\pm$12  \tablefootmark{c} & 3.0 & 54.5 & 0.1 \\
        13 And         &220885 &4.03$\pm$0.05\tablefootmark{i}   &1.87$\pm$0.2\tablefootmark{i}  &   -                                     &  1.479\tablefootmark{d} & 23 (1990--2013)\tablefootmark{d} & -63.4          \tablefootmark{d} & 2.7 & 306  & 0.6 \\
        V473 Tau       &30466  &4.045$\pm$0.011\tablefootmark{j} &1.64$\pm$0.15\tablefootmark{j} &   -                                     &  1.407\tablefootmark{j} & 47 (1963--2010)\tablefootmark{j} & -11$\pm$3      \tablefootmark{j} & 2.5 & 71.9 & 0.1 \\
        BS Cir         &125630 &3.966$\pm$0.014\tablefootmark{k} &1.72$\pm$0.10\tablefootmark{k} &   -                                     &  2.204\tablefootmark{c} & 39 (1975--2014)\tablefootmark{c} &  17.0$\pm$1.3  \tablefootmark{c} & 2.5 & 493  & 0.8 \\
        CS Vir         &125248 &3.992$\pm$0.013\tablefootmark{a} &1.50$\pm$0.08\tablefootmark{a} & $9.0^{+1.1}_{-1.3}$\tablefootmark{a}    &  9.300\tablefootmark{a} & 83 (1928--2011)\tablefootmark{l} &  66$\pm$8    \tablefootmark{l} & 2.2 & 184  & 0.2 \\
        \hline
    \end{tabular}}
    \tablefoot{Columns: $T_\mathrm{eff}$, $L$, and $B_\mathrm{pole}$ are the effective temperature, the luminosity, and the magnetic field strength at the magnetic pole, respectively. $T_\mathrm{baseline}$ is the baseline duration of the observations for rotational periods. $M_\mathrm{ini}$, $\tau$, and $\tau/\tau_\mathrm{MS}$ are the initial mass, the age, and the main-sequence fractional age, respectively, that are estimated theoretically via fitting to the location in the HR diagram.
	References:
        \tablefoottext{a}{\citet{Shultz22}}
        \tablefoottext{b}{\citet{Oksala15_sigOriE}}        
        \tablefoottext{c}{\citet{Mikulasek16}}
        \tablefoottext{d}{\citet{Pyper20}}
        \tablefoottext{e}{\citet{Shultz19_V913Sco}}
        \tablefoottext{f}{\citet{Adelman01}}
        \tablefoottext{g}{\citet{Mikulasek22}}
        \tablefoottext{h}{\citet{Kochukhov14}}
        \tablefoottext{i}{SIMBAD photometry}
        \tablefoottext{j}{\citet{Ozuyar18_V473Tau}}
        \tablefoottext{k}{\citet{Kochukhov06}}
        \tablefoottext{l}{\citet{Ozuyar18}}
    }
    \label{table-list}
\end{table*}

In this study, we analyzed ten magnetic stars that exhibit variations in their rotation periods. These stars are listed in Table \ref{table-list}, which displays their parameters, including surface effective temperature, luminosity, and the strength of the magnetic field at the pole. These values are collected from the literature that is indicated by the footnote of the table. As we could not find spectroscopic values for 13 And, a photometric estimate has been done using the B and V band magnitudes listed in SIMBAD (mB=5.726, mV=736) with a bolometric correction \citep{Allen00} and a parallax of 10.3618 mas. The uncertainty is roughly estimated based on the errors of photometric values relative to their spectroscopic values for other listed stars. Also, for V343 Pup, 13 And, V473 Tau, and BS Cir, we couldn't find literature indicating their surface magnetic field strength, so we assumed a typical value of 1 kG. 
To detect rotational period variations, it is essential to have extended, regular monitoring of the rotation period over the long term. Hence, the table includes the length of the monitoring period, $T_\mathrm{baseline}$, for each magnetic star. 
The time derivative of the rotation periods, which is the direct observational indication of the rotational period variation, is also shown.

An increase in the rotation period, indicating a deceleration in rotational velocity, was detected from 
$\sigma$ Ori E \citep{Townsend10, Mikulasek16}, 
V901 Ori \citep{Mikulasek08, Mikulasek11, Mikulasek16, Pyper20}, 
56 Ari \citep{Adelman91, Adelman01}, 
V343 Pupis \citep{Mikulasek22},
BS Cir \citep{Mikulasek16}, 
and CS Vir (\citealt{Ozuyar18}, but see \citealt{Pyper20}).
On the other hand, 
V913 Sco \citep{Shultz19_V913Sco, Pyper20}, 
13 And \citep{Pyper20},
and 
V473 Tau \citep{Ozuyar18_V473Tau}
exhibit a decrease in the rotation period, implying an acceleration in their rotational velocities.
Furthermore, CU Vir exhibits a transition in its rotational period variation from increasing to decreasing (\citealt{Adelman92, Pyper98, Mikulasek11, Mikulasek16}, but for alternative interpretation, see \citealt{Pyper20}), suggesting the possibility of cyclic nature of the variation. 
While different authors have reached different conclusions, the change of the sign of the rotational period variations may also occur for V901 Ori \citep{Mikulasek16} and V913 Sco \citep{Shultz19_V913Sco}. 
The cyclic changes in the rotational period variation take place on timescales of several decades, with estimated periods of 67.6 years for CU Vir, 60 years for V913 Sco, and more than 100 years for V901 Ori. 

\begin{figure}
    \centering
    \includegraphics[width=\linewidth]{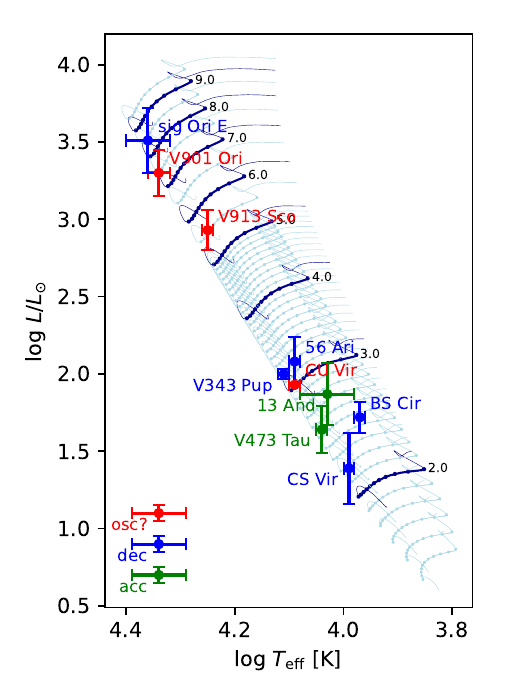}
    \caption{
The ten magnetic stars where rotational period variations have been detected are shown on the HR diagram. Those showing a long-term increase in rotation period are represented in blue (decelerating), those with a decrease in green (accelerating), and those transitioning from an increase to a decrease (or vice versa) in red (oscillating). The solid curves represent the results of evolution calculations of nonrotating, nonmagnetic stars. Models with ZAMS masses from 1.5 to 5 $M_\odot$ are shown at intervals of 0.1 $M_\odot$, while models with ZAMS masses from 5.5 to 9.5 $M_\odot$ are shown at intervals of 0.5 $M_\odot$. The numbers displayed next to the solid curves represent the ZAMS mass. To indicate the ages, points were placed at 10 equal divisions of the period from ZAMS to TAMS.
    }
    \label{fig-HR}
\end{figure}

We prepare a set of nonrotating, nonmagnetic stellar evolution simulations using the stellar evolution code HOSHI \citep{Takahashi&Langer21}. The results are compared to the magnetic stars on the HR diagram, which is shown as Fig.~\ref{fig-HR}. This allows us to select the best-fit models to determine each object's zero-age main-sequence (ZAMS) mass and age, which are indicated in the table. The stars' ages are presented both as mega-years and as fractional ages normalized to the duration of the main sequence stage.
While magnetic fields and rotation may exert considerable influences on tracks in the HR diagram \citep{Petermann+15,Keszthelyi+19, Keszthelyi20}, their effects are not essential for the scope of this study.

We searched the literature for indications of these stars belonging to clusters or associations.
$\sigma$ Ori E is thought to belong to the $\sigma$ Orionis Open Cluster \citep[e.g.,][]{Tarricq21_sigOri}, which is estimated to be 3-5 Myr old \citep{Osorio02_sigOri}. 
V901 Ori is suggested to belong to the Ori B star-forming region, which has an age of 0.8-2.2 Myr \citep{Roman-Zuniga23_OriB}, 
while V913 Sco may be part of Upper Sco with an age of 10-12 Myr \citep{Esplin18_UpperSco, Luhman20_UpperSco}. 
V343 Pup is also thought to belong to the Vela OB2 complex \citep{Cantat-Gaudin19_VelaOB2}. The age of this association is estimated to be 7.5-10 Myr from isochrones, which is somewhat inconsistent with the lithium depletion observed in pre-main-sequence stars \citep{Prisinzano16_GaiaESO_gamVel, Jeffries17_GaiESO_gamVel}.
While the estimated ages for the parent regions do not perfectly match our age estimates, no significant discrepancies are found due to the large uncertainties in our estimates.
We do not find works suggesting relationships with known clusters and associations for the other stars in our sample.

\subsection{Stellar models}

The HOSHI code we used to compute stellar models is a general-purpose stellar evolution code designed to handle standard physics such as radiative and convective energy transfer, time-dependent chemical mixing, nuclear reactions, as well as stellar rotation and stellar magnetism.
Here we briefly describe the most relevant elements of the code, the treatment of rotation and magnetic fields, and a more detailed description can be found in \citet{Takahashi&Langer21}.

The code uses angular velocity as one of the independent variable to treat stellar rotation. 
As a critical assumption, our formulation treats that the angular velocity of the gas can be expressed as a function of the enclosed mass $\Omega(M)$. This relies on the assumption known as the 'shellular rotation law' \citep{zahn92}, which adopts a picture where viscosity due to horizontal turbulence homogenizes the angular velocity on equipressure surfaces. The validity of this assumption will be discussed later.

The time evolution of the angular velocity distribution is determined by the following diffusion equation with an additional term that shows magnetic stress:
\begin{equation}
    \begin{split}
    \frac{d}{dt}(i\Omega)
    = & \frac{\partial}{\partial M}\left(
        (4\pi \rho r^2)^2 \nu_\mathrm{cv} i
        r^{-n_\mathrm{cv}}
        \frac{\partial (\Omega r^{n_\mathrm{cv}})}
        {\partial M}
        \right) \\
    & + \frac{\partial}{\partial M}\left(
        (4\pi \rho r^2)^2 \nu_\mathrm{eff} i
        \frac{\partial \Omega}
        {\partial M}
        \right)
    + \frac{\partial}{\partial M}\left(
        \frac{8 r^2 A B}{15}
        \right),
    \end{split} \label{eq-omega}
\end{equation}
where $i, \rho, r, A$, and $B$ are the specific moment of inertia, density, radius, toroidal vector potential expressing poloidal magnetic component, and toroidal magnetic component, respectively, all of which are functions of mass.
$\nu_\mathrm{cv}$ is the viscosity due to convective turbulence and is estimated as $\nu_\mathrm{cv} = v_\mathrm{cv} l_\mathrm{cv} / 3$ using the convective velocity $v_\mathrm{cv}$ and length scale $l_\mathrm{cv}$ provided by the mixing-length theory.
A fudge factor $n_\mathrm{cv}$ is introduced to absorb our ignorance of angular momentum transport due to convective turbulence. In this work, $n_\mathrm{cv} = 0$ is used so that convective regions rotate rigidly.
$\nu_\mathrm{eff}$ is the effective viscosity due to turbulence caused by instabilities other than convection.
The dynamical and secular shear instability, the Solberg-Hoiland instability, and the Goldreich-Schubert-Fricke instability are included here as instabilities due to rotation \citep{Pinsonneault+89, Heger+00}, and the Pitts-Tayler instability as instability due to magnetic fields \citep{Spruit02, Maeder&Meynet04}.
The boundary condition is to set all angular momentum fluxes to zero at the inner and outer boundaries.

The evolution of the stellar magnetic field is represented by the time evolution of two functions, $A(M)$ for the poloidal component and $B(M)$ for the toroidal component:
\begin{equation}
    \frac{d}{dt}(Ar)
    =  4 \pi \rho \eta r^3 \frac{\partial}{\partial M}\left(
       4\pi \rho 
       \frac{\partial (A r^2)}{\partial M}
       \right), \label{eq-poloidal}
\end{equation}
\begin{equation}
    \begin{split}
    \frac{d}{dt} \left( \frac{B}{\rho r} \right)
    = & 4 \pi A r \frac{\partial \Omega}{\partial M}
        + 4 \pi \eta r^2 \frac{\partial}{\partial M}\left(
         \frac{4\pi\rho}{r^2}
        \frac{\partial (B r^3)}{\partial M}
        \right) \\
      & + (4\pi)^2 \rho r^2 
            \frac{\partial \eta}{\partial M}
            \frac{\partial (Br)}{\partial M}.
    \end{split} \label{eq-toroidal}
\end{equation}
$\eta$ represents the effective magnetic viscosity. It is estimated as $\eta = \nu_\mathrm{cv} + f_\mathrm{m} \times \nu_\mathrm{eff}$ with a parameter $f_\mathrm{m}$.
The control parameter $f_\mathrm{m}$ is introduced to handle the large uncertainty of the turbulent transport coefficients, but $f_\mathrm{m}=1$ is used in this study.
The inner boundary conditions are chosen as
    $\partial(A/r)/\partial M = 0$
and 
    $\partial(B/r^2)/\partial M = 0$
so that the magnetic fields do not diverge and are smoothly connected to the solution of the diffusion equations at the center.
At the stellar surface, we assume that the poloidal component smoothly connects to the external dipole field, and the toroidal component is set to zero because the electric current is assumed not to penetrate the stellar surface.
These are expressed by $\partial(Ar^2)/\partial M=0$ and $B=0$.

\begin{figure*}
    \centering
    \includegraphics[width=\linewidth]{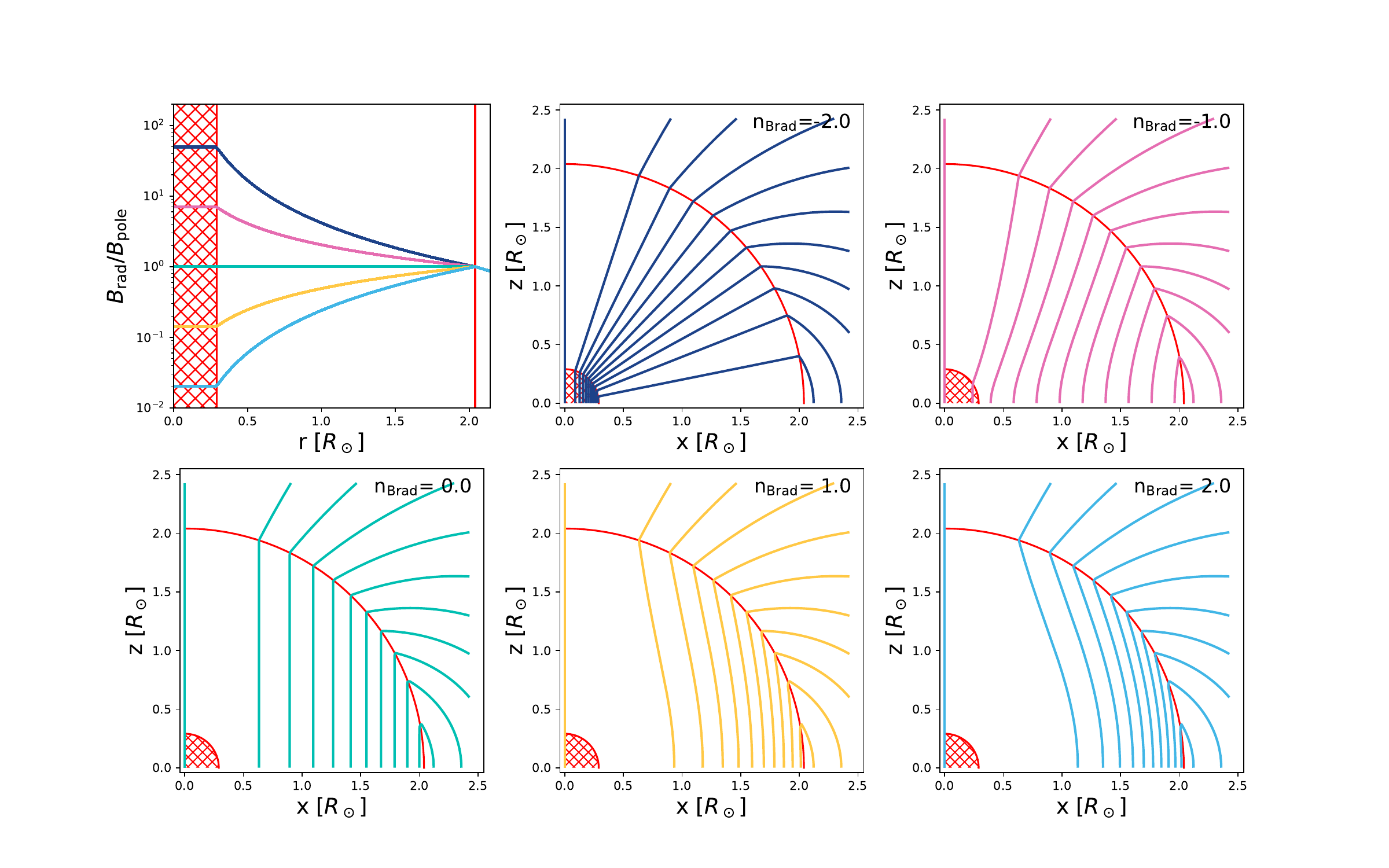}
    \caption{
    The examples of parametric poloidal field structures that are applied in this work.
    Solid lines in the top-left panel show the radial dependencies of $B_{\mathrm{rad}}/B_{\mathrm{pole}}(r)$ functions 
    with $\nb$ = -2.0 (dark blue), -1.0 (magenta), +0.0 (green), +1.0 (yellow), and +2.0 (cyan).
    The red-hatched region in the stellar center corresponds to the central convective region, 
    in which the magnetic field is assumed to be constant.
    The vertical red-solid line indicates the stellar surface, where $B_{\mathrm{rad}}/B_{\mathrm{pole}}(r) = 1$.
    The rest plots show the 2D poloidal field structures in the meridional cuts.
    As in the top-left panel, the red-hatched area shows the central convective region
    and the red-solid curve is the stellar surface.
    }
    \label{fig-2DBfield}
\end{figure*}

\subsection{The eigenmode analysis}

\citet{Takahashi&Langer21} showed that the eqs.~\ref{eq-omega} and \ref{eq-toroidal} can be reduced to a wave equation representing the propagation of torsional \Alfven{} waves if the time evolution of density, radius, poloidal magnetic component and other coefficients are negligible.
This means that the standing waves of the torsional \Alfven{} wave of interest in this study can be expressed as eigenmodes of this wave equation.
Therefore, we conduct an eigenmode analysis to find standing-wave solutions instead of directly solving the wave propagation and scanning the frequency space as done in \citet{Krticka+17}.
With this method, each eigenmode, including higher-order modes can be efficiently and robustly obtained.
It has been confirmed that the obtained eigenmodes describe oscillatory time evolution by actually solving the dynamical evolution for several models.

The independent variables of the wave equation are $\Omega(t, M)$ and $B(t, M)$, and it can be schematically expressed as
\begin{equation}
    \begin{split}
    \frac{\partial \Omega}{\partial t} &= F_\Omega (\Omega, B) \\
    \frac{\partial B}{\partial t} &= F_B (\Omega, B).
    \end{split} \label{eq-wave}
\end{equation}
Here, $F_\Omega$ and $F_B$ are terms expressing the advection and diffusion in eqs.~\ref{eq-omega} and \ref{eq-toroidal}. They are partial derivatives in the original equations, but in numerical calculations, they are expressed as linear functions of the $\Omega$ and $B$ vectors.
Hence, if we assume that the two independent variables have a common time dependence of 
$
    \Omega \sim B \sim \exp(\lambda t),
$
it is further reduced to a linear equation as
\begin{equation}
    \lambda \ \mathbf{x} = F \cdot \mathbf{x}, \label{eq-linear}
\end{equation}
where $\mathbf{x} = (\Omega, B)^t$ is a vector arranging the independent variables in a row, and $F$ is a matrix showing the advection and diffusion terms.
Eq.~\ref{eq-wave} indeed forms an eigenvalue problem of the matrix $F$ with the eigenvalue of $\lambda$ and the associated eigenvector of $\mathbf{x}$.

The eigenvalue $\lambda$ is decomposed into the real part $\sigma$ and the imaginary part $\omega$, 
\begin{equation}
    \lambda = \sigma + i \omega, \nonumber
\end{equation}
where $i$ in this equation shows the imaginary unit, and a non-zero imaginary part means that the eigenmode shows an oscillatory evolution. We interpret this oscillatory evolution as the rotational period variation observed in some magnetic stars. Hence, the period of the variation, $\Pi$, can be estimated as $\Pi = 2 \pi / \omega$ using the eigenvalue in this model. We will investigate the properties of $\Pi$ later on.
One of the limitation of this model is that, since it is based on a linear analysis, it cannot predict the amplitude of the rotational period variation, which requires some kind of driving mechanism that is beyond the scope of the present study.

In practice, we evaluate the matrix $F$ from partial derivatives of $F_\Omega$ and $F_B$ as
\[
    F = \begin{pmatrix}
            \frac{\partial F_\Omega}{\partial \Omega} & \frac{\partial F_\Omega}{\partial B} \\
            \frac{\partial F_B}{\partial \Omega} & \frac{\partial F_B}{\partial B} \\
        \end{pmatrix}
\]
since they are already prepared to form the Jacobian in the main Newton-Rhapson iteration for evolution simulations. 
Hence, our eigenmode analysis shares the spacial resolution with the stellar evolution simulation. The mesh number is typically $\sim 1000$.
Both $F_\Omega$ and $F_B$ actually include nonlinear terms of $\Omega$ and $B$ in the diffusion coefficients, which we assume to be negligible. 
This assumption seems reasonable, since the direct simulation of the time evolution with the obtained eigenmodes as input reproduced the standing waves very well. 
The eigenmode analysis of the $F$ matrix is done using the DGEEV subroutine in the open library LAPACK.

\subsubsection{The parametric structure of the poloidal magnetic field}
The magnetic field structure of stellar interiors remains a perplexing subject.
In this study, the strategy is to describe the internal structure of the poloidal magnetic field using a simple parametric function with minimal assumptions.
First, the magnetic field is assumed to be axisymmetric to the axis of rotation, as in the basic definition of our evolution simulation.
Second, the radial dependence is expressed using the parameter $\nb$ as
\begin{equation}
    B_\mathrm{rad}(r) = B_\mathrm{pole} \times \left\{
        \begin{array}{ll}
            (R_\mathrm{cv}/R_\mathrm{star})^{\nb} & (r<R_\mathrm{cv}) \\
            (r            /R_\mathrm{star})^{\nb} & (r>R_\mathrm{cv}),
        \end{array}
    \right.
\end{equation}
where $R_\mathrm{cv}$ and $R_\mathrm{star}$ are the radius of the central convective region and the stellar radius and $B_\mathrm{pole}$ is the magnetic flux density at the pole on the stellar surface. 
The magnetic stars considered in this work are early-type main-sequence stars, and all of them have a convective region in the center. Therefore, the magnetic field does not diverge or go to zero at the center using this expression.
Then, the radial field is converted into the $A$ function as $A = B_\mathrm{rad} \times r/2$.
Figure~\ref{fig-2DBfield} shows magnetic field structures with different $\nb$ applied to the CU Vir model as an example.

\subsubsection{A correlation between $\Pi$ and $B_\mathrm{pole}$}
As noted in \citet{Krticka+17}, the eigenvalue $\lambda$ follows a scaling transformation of $\lambda' = \gamma^{-1} \lambda$ with $B_\mathrm{pole}' = \gamma B_\mathrm{pole}$ due to the linear dependence of the matrix $F$ on $B_\mathrm{pole}$.
Hence, a linear relationship between $\Pi$ and $B_{\rm{pole}}$ can be obtained:
\begin{equation}
	\Pi \times B_{\rm{pole}} = const. \label{eq-correlation}
\end{equation}
One may calculate $\Pi$ for models with different values of $B_{\rm{pole}}$ using this relationship. Furthermore, the uncertainty in $B_\mathrm{pole}$ is associated with the uncertainty in $\Pi$ through this relationship.

\section{Results} \label{sec:Results}

\subsection{General trends of the eigenmode analysis}

\begin{figure}
    \centering
    \includegraphics[width=\linewidth]{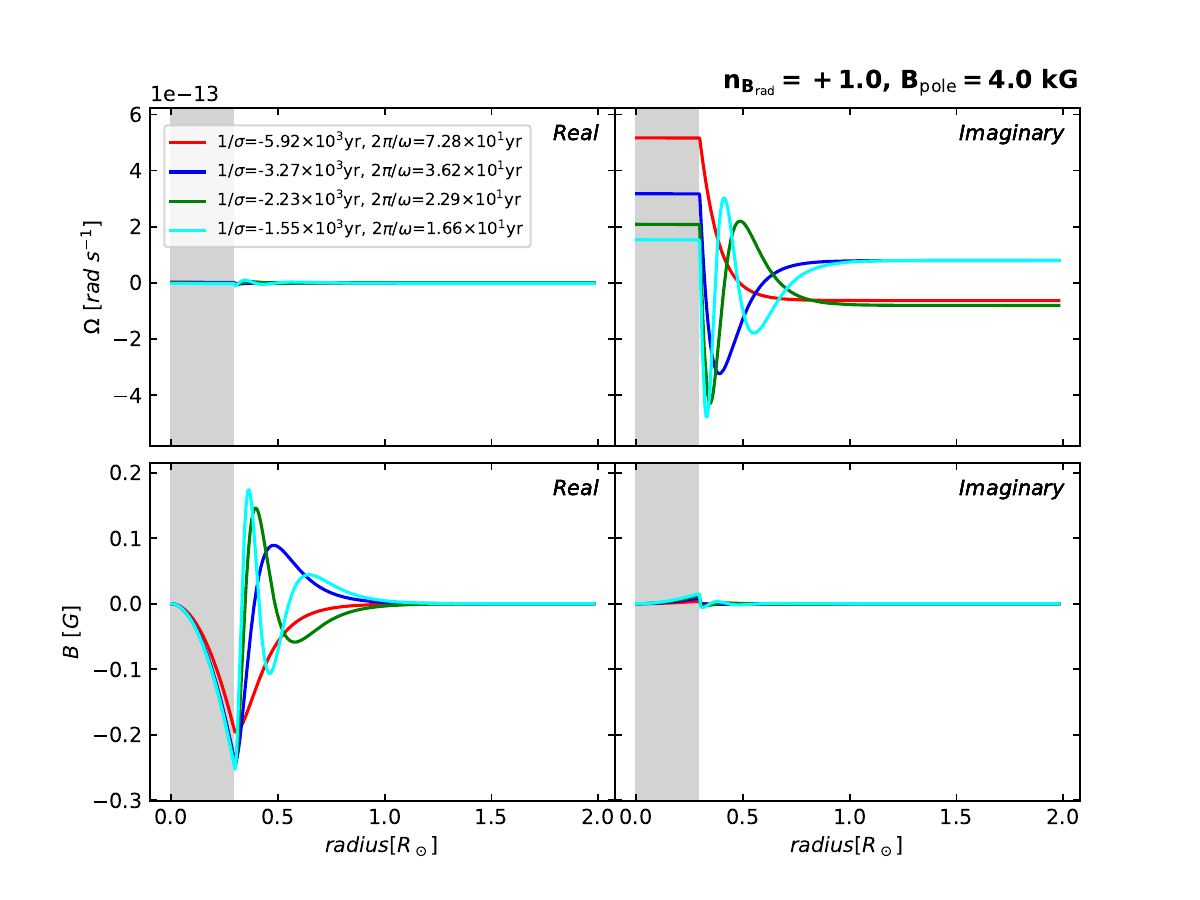}
    \caption{
    Eigenfunctions obtained for the CU Vir model with $\nb=+1.0$.
    The top two panels display the distribution of angular velocity, $\Omega(r)$, 
    while the bottom two panels show the toroidal magnetic field component, $B(r)$. 
    In each row, the left panel shows the real part, and the right panel shows the imaginary part.
    The modes from the lowest to the fourth are indicated as red, blue, green, and cyan lines.
    The reciprocals of the eigenvalues for each mode are shown in the legend.
    The gray-shaded area indicates the central convective region.
    }
    \label{fig-check_eigen}
\end{figure}

First, we discuss the general trends of the eigenmode analysis by taking the CU Vir model as an example.
The background structure is taken from a model with $M_\mathrm{ini}=$ 3.0 $M_\odot$ and $\tau=$ 54.5 Myr.
In Fig.~\ref{fig-check_eigen}, lowest-order eigenmodes from the first to the fourth obtained for the model with $\nb=+1.0$ are shown as a function of radius. In the central convective region, $\Omega$ profiles become flat and $B(r)$ profiles become $\propto r^2$ due to the large diffusion coefficient. Outside the convective region, both profiles show characteristic oscillatory patterns, whose number of nodes corresponds to the order. For example, for the lowest order mode, the number of nodes in the $\Omega$ and $B$ profiles are 1 and 0, respectively, and it increases by one as the order of the eigenmode increases. And the frequency of the torsional oscillation, $\Pi = 2 \pi / \omega$, decreases as the order of the mode increases.

\begin{figure}
    \centering
    \includegraphics[width=\linewidth]{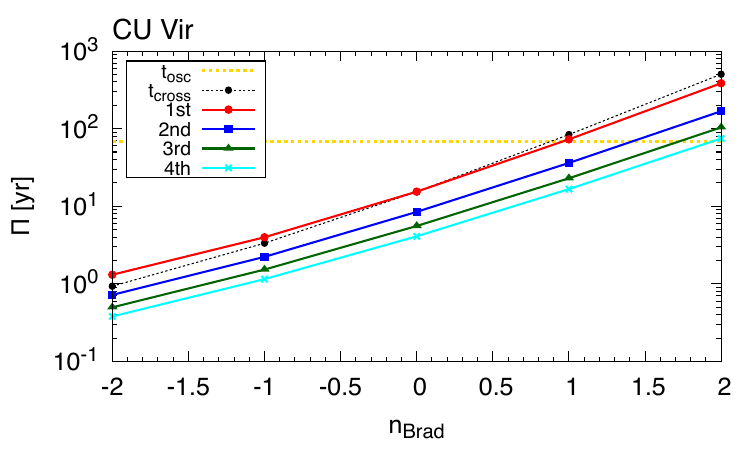}
    \caption{
The $\nb$ dependence of oscillation periods estimated for the CU Vir model is shown for eigenfunctions from the first to the fourth orders. Different line styles are used for first (red, circle), second (blue, square), third (green, triangle), and fourth (cyan, cross). For comparison, $t_{\rm cross}$ is indicated by a black dotted line. The constant value shown as a yellow dashed line corresponds to the oscillation period suggested for CU Vir (67.6 yr).
    }
    \label{fig-eigen_values_CUVir}
\end{figure}

The $\nb$ dependence of the period of torsional oscillation is shown in Fig.~\ref{fig-eigen_values_CUVir}. 
With the observational constraint of $B_\mathrm{pole} = 4.0 \ \mathrm{kG}$ for CU Vir \citep{Kochukhov14, Shultz22}, $\nb$ becomes the only parameter left to specify the internal poloidal field structure in our formulation.
The figure shows that the larger the value of $\nb$, the longer the oscillation periods irrespective to the order. Hence, there is a simple one-to-one relationship between the parameter $\nb$ and the frequency of oscillations. This relationship indicates that, within a crude approximation of the internal B-field structure, the structure of the internal poloidal field can be inferred by observing the period of torsional oscillations. Specifically, there are long enough observations for CU Vir, indicating that the rotation period may oscillate with a period $\Pi$ = 67.6 yr. This period can be explained as the fundamental mode of the torsional oscillation with $\nb \sim +1.0$, indicating that the internal B-field is slightly concentrated on the stellar surface, rather than the center, for this particular case.

The figure also shows that the frequency of the lowest-order, which is shown by the red solid line, agrees well with the crossing time of the torsional \Alfven{} wave, the black dotted line.
The crossing time is estimated as
\begin{equation}
    t_\mathrm{cross} = \int_0^{R_\mathrm{star}} \frac{dr}{c(r)},
\end{equation}
where $c = B_\mathrm{rad}/\sqrt{20 \pi \rho}$ is the characteristic speed of the torsional \Alfven{} wave in our formulation \citep{Takahashi&Langer21}. 
This equation indicates that the region with the slowest wave propagation speed dominates the overall crossing time. Additionally, the strength of the poloidal magnetic field is proportional to the wave propagation speed and inversely proportional to the crossing time. This gives a physical explanation of the relation eq.~(\ref{eq-correlation}).

\subsection{Variation periods with different background structures}

\begin{figure}
    \centering
    \includegraphics[width=\linewidth]{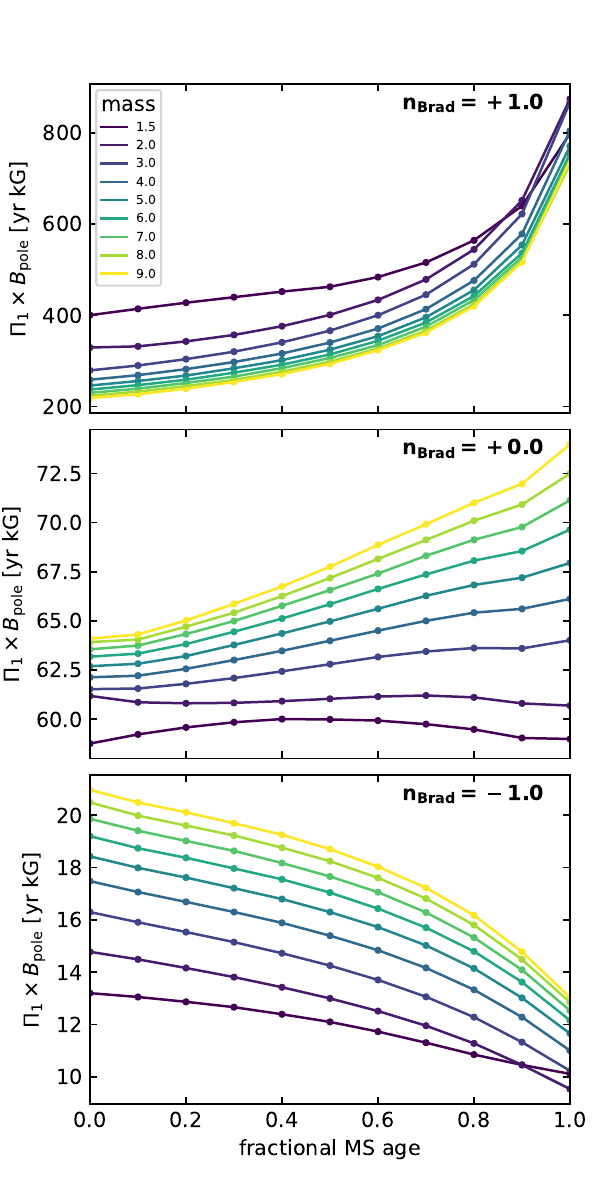}
    \caption{Variation periods $\Pi$ for different background structures with different fractional main-sequence ages and masses. 
    Three panels show results with different $\nb$ of +1.0 (top), +0.0 (middle), and -1.0 (bottom).
    The dependency on $B_\mathrm{pole}$ is implicitly included in the y-axis based on the relation eq.(\ref{eq-correlation}).
    }
    \label{fig-mass_dependence}
\end{figure}

The variation period $\Pi$ depends not only on the magnetic field structure but also on the background density structure. 
Figure \ref{fig-mass_dependence} illustrates how $\Pi$ varies as the mass and age of the background structure model are changed while keeping the parameters specifying the magnetic field structure.
It not only shows that $\Pi$ depends on the background model, but also demonstrates how this dependency varies with $\nb$.
Namely, for the case where the magnetic field is concentrated on the surface ($\nb = +1.0$), $\Pi$ increases with increasing age and decreasing mass. 
Conversely, for the case where the magnetic field is centrally concentrated ($\nb = +1.0$), 
the trend reverses, and $\Pi$ increases with decreasing age and increasing mass. 
However, the change in $\Pi$ due to differences in background structure remains at most a few times, which is relatively small compared to the change due to differences in $\nb$, which can span several orders of magnitude.

\begin{figure}
    \centering
    \includegraphics[width=\linewidth]{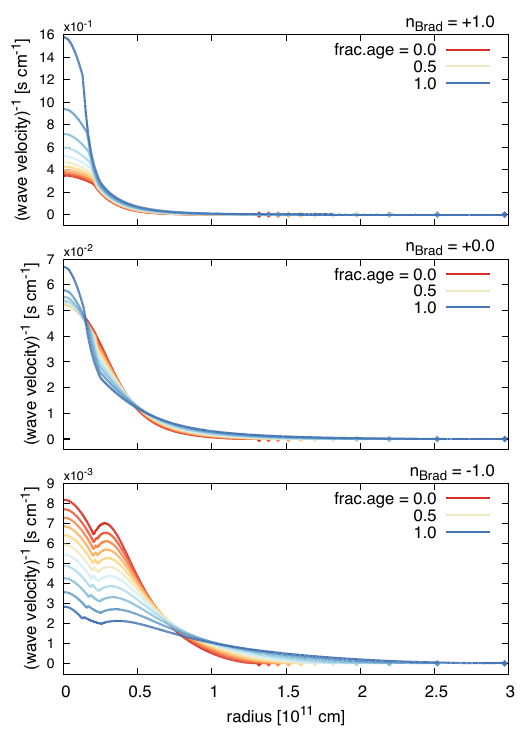}
    \caption{
    The inverse of the velocity of torsional \Alfven{} wave is shown as a function of radius such that the enclosed area by the curves corresponds to the crossing time. 
    From top to bottom, results for $\nb=+1.0, +0.0$, and $-1.0$ with $B_{\rm pole} = 1$ kG are shown.
    The density distributions are based on the 3 $M_\odot$ model. Each panel includes 11 lines corresponding to different fractional MS ages of 0.0, 0.1, …, 1.0, with the color of the lines changing from red to blue.
    Filled circles represent a change in radius.
    }
    \label{fig-structure}
\end{figure}

Different values of $\nb$ lead to varying dependencies on the background, and most importantly, changing the value of $\Pi$ significantly. This characteristic should be closely related to the velocity distribution of torsional \Alfven{} waves within the star, in particular, the distribution of propagation time in bottleneck regions.
To explore this, the distribution of inverse of the \Alfven{} velocity is plotted in Fig.~\ref{fig-structure} for $M_\mathrm{ini}=3 M_\odot$ models with different ages. The top panel illustrates the results for $\nb = +1.0$, while the middle and bottom panels are for $\nb = +0.0$ and -1.0, respectively. In this figure, the area enclosed by the lines and the x-axis corresponds to the crossing time.

The figure shows that the region where \Alfven{} waves slow down the most is the central part of the star, regardless of the magnetic field structure. This indicates that, even with $\nb=-1.0$, the increase in density at the center is responsible for determining the speed of \Alfven{} waves. On the other hand, despite keeping the surface magnetic field strength constant, the speed of \Alfven{} waves changes by an order of magnitude with different $\nb$. This is why the crossing time and $\Pi$ show the large $\nb$ dependence.

In the case of $\nb=+0.0$, the wave velocity distribution remains relatively unchanged as the star ages. 
In this model, a uniform magnetic field of constant strength is imposed throughout the stellar interior, and the wave velocity is only affected by variations in the density distribution. Therefore, the time-constant nature indicates that changes in the density distribution during main-sequence star evolution itself do not have a significant impact on the oscillation period evolution.
On the other hand, for $\nb=\pm1.0$, the changes in density distribution have a stronger impact on the variation period. This is because, when using our magnetic field structure description, the field strength in the central region directly depends on the stellar radius. The larger the age of the model, the larger the radius. Hence, in the case of $\nb=+1.0$, the central magnetic field becomes weaker, causing the velocity to slow down. Conversely, for $\nb=-1.0$, the central field becomes stronger, resulting in faster wave velocities.
In other words, the change in the variation period with different ages can be attributed more to the simplification of the magnetic field structure used in this work rather than being a direct physical consequence of stellar evolution.

\subsection{Eigenmode analysis for specific magnetic stars}

\begin{figure*}
    \centering
    \includegraphics[width=\linewidth]{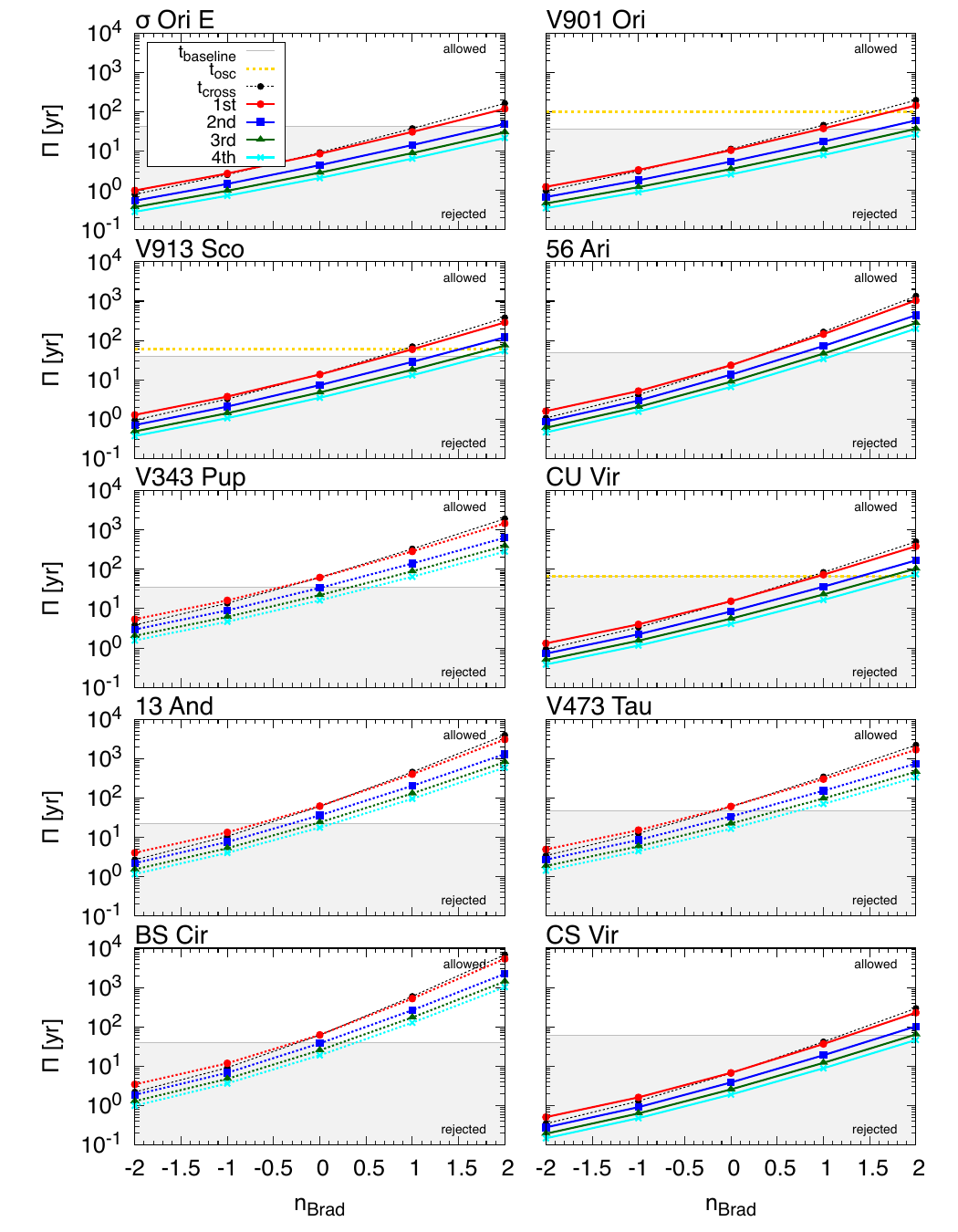}
    \caption{
The same as Fig.~\ref{fig-eigen_values_CUVir}, but for all the magnetic stars where rotational period variations have been observed.
    }
    \label{fig-eigen_values}
\end{figure*}

The same analysis as conducted on CU Vir is applied to the other nine magnetic stars that are listed in Table 1, and the results are shown in Fig.~\ref{fig-eigen_values} for all ten stars. To perform eigenmode analysis using our model, it is necessary to set the magnetic field strength on the stellar surface. As noted in section 2., we couldn't find such values for V343 Pup, 13 And, V473 Tau, and BS Cir, so a typical value of $B_\mathrm{pole}$ = 1 kG is applied for these stars. The obtained periods follow the relationship $\Pi \times B_\mathrm{pole} = const$. Consequently, they may linearly shift up and down based on the actual field strength. To indicate this additional uncertainty, the results of the eigenvalues for the two stars are shown in dashed lines.

Detecting rotational period variations requires long-term monitoring, and for each star, the duration of this extended monitoring is indicated as $T_\mathrm{baseline}$. It is expected that modes with periods shorter than $T_\mathrm{baseline}$ have already been detected through previous monitoring. However, this contradicts the fact that there have been no cases where the periodicity of rotational period variations has been definitively confirmed, even for CU Vir. Consequently, magnetic field structures that have such modes can be rejected. On the other hand, field structures with longer periods than $T_\mathrm{baseline}$ are allowed observationally. In the figure, this condition is represented with a gray or white background. Based on this condition, it is concluded that centrally concentrated magnetic field structures with a negative $\nb$ is rejected for all magnetic stars.

Although not yet settled, changes in the sign of rotational period variations have been detected for V901 Ori, V913 Sco, and CU Vir, and these changes allow for the estimation of the variation periods. These values are represented by yellow horizontal dashed lines in the figure. Our theoretical model, with its strong dependency on $\nb$, allows for explanations of a wide range of the value. As a consequence, the periods for these stars can be explained by setting $\nb \gtrsim 1.0$. This might suggest that the magnetic field structure inside the stars is strong near the surface.

\section{Discussion} \label{sec:Discussion}

\subsection{Limitations of our model}
In our framework, a key assumption for one-dimensionalization is the shellular rotation law, where the material at the same radius (originally on an isopressure surface) is assumed to have the same angular velocity. This assumption may be justified if structural changes occur on the timescale of stellar evolution, since it is sufficiently long, allowing turbulent viscosity originating from horizontal turbulence to relax the angular momentum distribution. However, it is uncertain whether the same process occurs on the \Alfven{} time that is much shorter than the evolutionary timescale. Instead, one might consider that differential rotation develops in the horizontal direction during the propagation of \Alfven{} waves.

Without assuming shellular rotation, \Alfven{} waves traveling along different magnetic field lines would have distinct crossing times. In this case, differential rotation between gas associated with adjacent magnetic field lines may induce turbulence, causing phase mixing \citep[cf,][]{Charbonneau92_AngMom_MHD, Charbonneau93_AngMom_MHD}.
To determine oscillation periods of stationary \Alfven{} waves in this context, a two-dimensional linear analysis, considering the efficiency of phase mixing, will be suitable.

Taking this into account, in the current one-dimensional framework, the model's accuracy is expected to decrease as the crossing time for each magnetic field line becomes different. Figure 2 suggests that in centrally concentrated magnetic field structures ($n \lesssim 0$), where path lengths are close, the model's accuracy might not be severely affected. Conversely, in surface-concentrated magnetic field structures ($n \gtrsim 0$), there could be a potential mismatch between the predictions of our one-dimensional model and future two-dimensional models.

Another influential simplification is our assumption that the rotation and the magnetic axes are aligned. In reality, it is common for the two axes not to align, so that analyses using the Oblique Rotator Model are commonly conducted.
However, our analysis has also shown that the period of rotational period variations is simply determined by the \Alfven{} wave crossing time. Therefore, even when considering the obliquity, we expect only a few-fold difference in the variation periods, which is unlikely to affect the conclusions of this study.

In upper main sequence stars, a strong dynamo-generated magnetic field may exist within the convective core \citep[e.g.,][]{Featherstone09_CoreDynamo_FossileField, Augustson16}. Dynamo-driven fields may show reversals of their polarity. It is thus reasonable to assume that the dynamo-generated field is not directly connected to the stable magnetic field in the radiative layer.
In a one-dimensional model, this disconnection may be represented by setting the magnetic field strength to zero at the boundary of the convective region. However, in our approach, the zero magnetic field causes the local crossing time to diverge, making it complicated to estimate the crossing time. Due to this issue, applying our model to cases with strong dynamo fields remains challenging.
However, the radius of the convective core region is relatively small, and its crossing time is not expected to dominate over the crossing time of the outer radiative layers. Therefore, our current analysis is expected to provide a reasonable order of magnitude estimate of the oscillation period.

\subsection{Comparison with Krticka's model}
\citet{Krticka+17} also analyzed the rotational period variations of CU Vir and V901 Ori based on the picture that the variations are induced by standing \Alfven{} waves. The one-dimensional magnetohydrodynamic equations they obtained are similar to ours, but they use a different poloidal magnetic field configuration that is expressed in the ($R, Z$) plane as
\begin{eqnarray}
	(B_R, B_Z) = \left( B, -\frac{zB}{R} \right),
\end{eqnarray}
where $B$ is a constant.
Their model yielded oscillation periods of 51 years for CU Vir and 5 years for V901 Ori. As a result, they concluded that the rotational period variation observed in CU Vir, with a period of approximately 70 years, can be explained by standing \Alfven{} waves. However, the variation in V901 Ori, estimated to have a period of over 100 years, cannot.

While dealing with similar models, our work has resulted in different conclusions. The main factor contributing to this difference is likely to be the different assumptions about the poloidal magnetic field structure. As demonstrated earlier, our model can alter the period of standing Alfven waves by several orders of magnitude by changing the radial dependence of the poloidal magnetic field. On the other hand, it can be inferred that Krticka's model, which has only tested a single field structure, had a limited range of explainable periods.
Regarding the surface magnetic field, Krticka applied $B_{\mathrm{p}} = 10$ kG, given the highly complex surface magnetic field structure. In contrast, we employed a dipole component of 6.1 kG obtained from \citet{Shultz19_rotation}. While using a weaker value results in longer oscillation periods, the effect is considered minor, at most doubling the period.

\subsection{Estimating the magnetic field structure based on the rotational period variation}
Centrally concentrated magnetic fields appear unlikely based on the lack of detected oscillations in most stars.
If they were, \Alfven{} waves would rather quickly pass through the entire star, resulting in a sufficiently short period of the rotational period variation to be detected.
Comparison with three stars that potentially show cyclic variations (V901 Ori, CU Vir, and V913 Sco) suggests that the degree of surface concentration of the internal poloidal field is moderate ($\nb \sim +1$).

The geometry of magnetic fields in equilibrium in the stellar interior has been investigated with MHD simulation \citep[e.g.,][]{Braithwaite&Spruit04, Becerra22} and with analytical approaches \citep[e.g.,][]{Lyutikov10, Duez&Mathis10, Akguen13}. The surface-concentrated magnetic field structures as described in our model ($\nb > 0$) resemble the equilibrium structures obtained numerically \citep{Braithwaite&Spruit04} more than the central-concentrated structure ($\nb < 0$).

If the estimates of surface concentration for the three stars that show cyclic variations were quantitatively accurate, the most massive of them, V901 Ori, exhibits the strongest surface concentration ($\nb\sim+1.7$), suggesting that the degree of surface concentration may correlate with the mass of the stars.
Another possibility is that, considering that the age of V901 Ori is almost zero while the fractional MS ages of V913 Sco and CU Vir are in the range of 0.1--0.3, the initially more surface-concentrated magnetic field would permeate deeper into the star as it evolves.
There are indications that the magnetic field structure of magnetic stars becomes simplified with age \citep{Rusomarov15, Silvester15}. It is also possible that such global evolution is occurring simultaneously within the stellar interior.
However, this evolutionary scenario may contradict the results of \citet{Braithwaite&Nordlund06}, where the poloidal field stabilized by the toroidal field rises to the surface due to magnetic diffusion over time, weakening the magnetic field strength near the center.

Our analysis is limited by the small sample size and the simple quantification methods. Additionally, V901 Ori is known to have a highly complex surface magnetic field structure \citep{Kochukhov11}, unlike the largely dipolar magnetic fields of CU Vir \citep{Kochukhov14} and V913 Sco \citep{Shultz19_V913Sco}.
Further research is necessary to validate trends and dependencies related to the internal magnetic field.

\subsection{Implications from strong magnetic fields detected in red giant stars}

\citet{Li23} reported that the shift in dipole mode frequency observed in red giants can be explained by a strong magnetic field of 20--150 kG above the H-burning shell. Applying the magnetic flux conservation to the radial distribution obtained from our 1.5 M$_\odot$ model yields a magnetic field strength of approximately 0.5--5 kG at a mass coordinate of $\sim$0.2 M$_\odot$ at the TAMS stage. Since $\sim$0.2 M$_\odot$ is larger than the maximum mass of the convective core ($\sim$0.12 M$_\odot$), this field strength can be interpreted as representing the strength of a fossil magnetic field in the radiative layer.
Most of the stars in Li’s sample were likely normal non-magnetic stars during their main sequence phase, so their surface magnetic field was at most on the order of $\sim$1 G. This means that the red giants analyzed by Li et al. had a centrally concentrated magnetic field during their main sequence phase, which contradicts our conclusions.

Li et al. provided an alternative interpretation for the origin of the frequency shift. Specifically, the kernel function used to estimate the frequency shift is also sensitive to the region inside the H-burning shell. The observed frequency shift can then be explained by a magnetic field of 200--300 kG in the former convective core region, which can be accounted for by a convective dynamo driven by hydrogen burning. 
Our results are more consistent with this latter interpretation. Hence, they lead to the conclusion that a convective dynamo operates during hydrogen burning, and the magnetic field it generates remains after the cease of the core convection and causes the frequency shift observed in red giants. 
It should be noted that the method we used in this study is agnostic regarding the magnetic field distribution inside the stellar convective core. Thus, we do not rule out the possibility that a strong magnetic field, which is likely disconnected from the radiative layer, exists within the convective core.

Main sequence stars with masses less than around 1 $M_\odot$ are considered to form radiative cores \citep{Kippenhahn&Weigert90}.
Even in such low-mass stars, it is predicted that a small convective core forms due to non-equilibrium hydrogen burning mediated by $^{3}$He and $^{12}$C occurring just before the star enters the ZAMS. However, it is uncertain whether such a short-lived convective core can efficiently drive a dynamo, and if it can, whether the magnetic field will survive without dissipating during the long main sequence phase that follows.
Hence, it is less likely that low-mass red giants possess strong internal magnetic fields similar to those of higher-mass red giants.

Therefore, the most direct way to determine the origin of the frequency shift will be to investigate whether similar shifts are seen in the low-mass red giants.
Considering the inverse correlation between age and mass, finding and analyzing low-mass red giants may be less feasible, and the least massive star in the sample by Li et al. is estimated to have a mass of about 1.04 solar masses. Nevertheless, the Kepler targets include a non-negligible number of low-mass red giants \citep[e.g.,][]{Yu18_KeplerRG}. Investigating older, less massive red giants could shed light on this question.

\subsection{Excitation mechanisms of standing \Alfven{} waves in magnetic AB stars}

Magneto-sonic oscillations are thought to exit in magnetars, as signified by the quasi-periodic oscillations in the tail of giant X-ray flares \citep[e.g.,][]{Strohmayer05_NS_Xrayoscillation}, as the oscillation frequency is well reproduced by global MHD simulations \citep{Glampedakis06_NS_oscillation, Gabler18_NS_oscillation}.
In this case, fractures in the crust of the magnetar are thought to disturb the hydrostatic equilibrium, which can provide the excitation of the MHD waves.

It is not known what could excite \Alfven{} waves in magnetic main sequence stars. Since the ages of the investigated magnetic stars span several to a few hundred Myrs,
which corresponds to $10^5\dots 10^7$ oscillation periods, it appears unlikely that an event relating to the formation of these stars excited oscillations that are still ongoing.
On the other hand, the nuclear evolution of the considered main sequence stars imposes changes in the rotation period of the order of $10^{-10}\,$s\,s$^{-1}$, which is comparable to the observed period variations of $10^{-2}\,$s\,yr$^{-1}$ $\sim$ $3 \times 10^{-10}\,$s\,s$^{-1}$. Given these rough estimates, it may appear possible that the observed period changes are produced by standing \Alfven{} waves driven by changes in the internal shear layers on the nuclear timescale.

Furthermore, coupling between the core convection and the envelope magnetic field could excite magneto-sonic waves. Indeed, for red giant stars, \citet{Lecoanet17_wave_conversion} found that internal gravity waves can be efficiently converted into magnetic waves under certain conditions. Finally, \citet{Krticka+17} suggested that coupling of a stellar wind with the stellar magnetosphere could excite \Alfven{} waves. While sufficiently strong winds are not predicted for late\,B and   A\,type stars, their presence is suggested for CU\,Vir based on its periodic radio emission \citep{Trigilio00_CUVir_Radio}.

\subsection{Towards future detections of rotational period variations}

If we assume that the typical surface concentration of the magnetic field resembles our case $\nb\sim+1$, the period of cyclic variations is expected to inversely correlate with the strength of the surface magnetic field (eq.\ref{eq-correlation}) and to show an increasing correlation with age and a decreasing correlation with mass (Fig.~\ref{fig-mass_dependence}). 
Based on these trends, it would be possible to devise strategies to discover new stars exhibiting rotational period variations, and even more, cyclic variations.

Among the ten magnetic stars analyzed in this study, most of those with known surface field strength have observational baselines comparable to the predicted periods for the case $\nb=+1$. Within the next few decades, it may be possible that the second time derivative of the rotation period will be measured in these stars.
56 Ari has an exceptionally short observational baseline compared to the $\nb=+1$ period, which may suggest that the rotational period variations of this star are not due to the \Alfven{} wave but rather due to precession \citep{Shore76, Adelman91}. 
Determinations of the rotational period for this star have not been reported since 2001 to our knowledge. The passage of more than 20 years will aid in the detection of the second derivative component of rotational period variations, which is essential for constraining models.

For stars without constraints on surface magnetic fields, we estimate the $\nb=+1$ periods for a typical field strength of 1 kG, and their values are generally one order of magnitude larger than observational baselines.
The difference in timescales suggests that these stars may have relatively strong surface magnetic fields, around 10 kG. Alternatively, they could possess relatively centrally concentrated internal magnetic fields ($\nb\sim+0$).
If the surface magnetic fields of these stars were weaker, around 100 G, it might contradict the idea that rotational period variations are due to \Alfven{} waves, suggesting instead that they are caused by precession.
Hence, when magnetic stars exhibiting variations of the rotational period are discovered, it is crucial to estimate the surface magnetic field strength through spectropolarimetric observations.

Until now, the detection of rotational period variations has been biased towards rapidly rotating stars with rotation periods of around one day. This is because a shorter rotation period makes it easier to meet the condition of observing as many cycles of rotation as possible, which is required for detecting rotational period variations \citep{Pyper20}.
However, such rapidly rotating magnetic stars are biased towards younger stars \citep{Fossati16_BFieldDecay, Shultz18, Keszthelyi20}, and it is known that the majority of magnetic stars have longer rotation periods, peaking around three days \citep{Kochukhov06, Netopil17}. 
It is important to search for rotational period variations in the more common magnetic stars with longer rotation periods and relatively older ages. 

Another factor, the amplitude, is preferable to be large for the detection of rotational period variation. In this study, linear analysis is conducted so that the investigation of the amplitude is not feasible. To do so, further research is needed on how the rotational priod variation is driven and to what extent the long-term growth and dissipation occur.

\section{Concluding remarks} \label{sec:Conclusion}

In this paper, we explore the existence and properties of torsional \Alfven{} waves in magnetic stars on the upper main sequence. Based on the concepts developed in \citet{Takahashi&Langer21}, we perform an eigenmode analysis of standing \Alfven{} waves in such stars and derive the most likely oscillation frequencies. We do this while assuming different degrees of the central- or surface-concentration of the magnetic field, which results in largely different frequencies.

We compare our results with the rate of change of the stellar rotation period, and find, as \citet{Krticka+17}, that they can be compatible with the periods of the standing \Alfven{} waves. However, our results imply compatibility only for the case that the magnetic fields in these stars are concentrated towards the stellar surface, and outline the potential of this type of technique to explore the properties of the magnetic field in stellar envelopes.

\begin{acknowledgements}
K.T. thanks to Kengo Tomida and Fabian Schneider for fruitful discussions.
This work was supported in part by JSPS KAKENHI Grant Numbers JP22K20377, JP24K17102.
\end{acknowledgements}

\bibliographystyle{aa} % style aa.bst
\bibliography{biblio}
%-------------------------------------------------------------------

%\begin{appendix} 
%\section{Appendix A}
%\end{appendix}

\end{document}